\documentclass[apjl]{openjournal}

\usepackage{xcolor}
\usepackage{textgreek}
\usepackage[utf8]{inputenc}
\usepackage[english]{babel}

\usepackage{hyperref}
\hypersetup{
    unicode, 
    colorlinks=true,
    linkcolor=linkcolor,
    citecolor=linkcolor,
    filecolor=linkcolor,
    urlcolor=linkcolor,
}
\usepackage{color,colortbl}
\definecolor{linkcolor}{rgb}{0.0,0.3,0.5}
\usepackage{tensind}
\tensordelimiter{?}
\DeclareGraphicsExtensions{.bmp,.png,.jpg,.pdf}
\usepackage{verbatim}
\usepackage[normalem]{ulem}
\usepackage{orcidlink}
\usepackage{soul}

\begin{document}
\title{Comments on ``Little ado about everything'' by A.\ Lapi et al. and on\\cosmological back-reaction}

\author{Julian Adamek \orcidlink{0000-0002-0723-6740}}
\affiliation{Institut f\"ur Astrophysik, Universit\"at Z\"urich, Winterthurerstrasse 190, 8057 Z\"urich, Switzerland}
\affiliation{D\'epartement de Physique Th\'eorique, Universit\'e de Gen\`eve, 24 quai Ernest-Ansermet, 1211~Gen\`eve~4, Switzerland}
\email{julian.adamek@uzh.ch}

\begin{abstract}
In two papers, A.\ Lapi et al.\ introduce and discuss what they call the $\eta$CDM model, a stochastic framework in which they claim that fluctuations in the density field at the scale of tens of Mpc due to structure formation would effectively drive the accelerated expansion of the Universe. They claim that this qualitative behaviour would emerge from the dynamics of standard cold dark matter alone, without introducing any new physics. In this short comment, I argue that such a proposition is implausible. Some of my remarks are relevant more generally to frameworks that try to describe cosmological back-reaction.
\end{abstract}

\maketitle

\section{Introduction}
\label{sec:intro}

The $\eta$CDM model, introduced and discussed by \citet{Lapi:2023plb,Lapi:2025mgr}, falls into the category of back-reaction models. Generally speaking, cosmological back-reaction is concerned with the question of how the large inhomogeneities that exist in the real Universe affect the description of the Universe as a whole. For example, one can ask to what extent the cosmological parameters inferred from observations are `dressed' by the presence of inhomogeneities. The topic is a vast one, but for the purpose of this comment we can narrow it down to a very specific type of back-reaction: the idea that the gravitational dynamics of a spacetime filled with standard cold dark matter alone could somehow lead to accelerated expansion. Importantly, in this type of model, acceleration would not occur in the homogeneous limit of the model and would only appear in the presence of inhomogeneities. I emphasise that the fundamental laws of physics (classical matter evolving under standard gravity without other interactions) are assumed unchanged throughout this discussion and that the different phenomenology would be entirely due to the state of the system (homogeneous versus strongly inhomogeneous).

In the specific implementation provided by the $\eta$CDM model, this is achieved roughly in the following way. The Universe is considered as an ensemble of patches sized in the range\footnote{\cite{Lapi:2023plb} state 30-50 $h^{-1}$\,Mpc while \cite{Lapi:2025mgr} state 10-50 $h^{-1}$\,Mpc.} of tens of Mpc. Each patch is then treated as a separate Friedmann model, but with a stochastic noise term added to the continuity equations of matter and radiation. The inclusion of radiation in the modelling seems largely irrelevant, so I will not discuss it here in detail. According to \cite{Lapi:2023plb}, the ``noise term [\ldots] subtends deviations from local energy conservation in different patches of the Universe with a size given by the aforementioned coarse-graining scale that can be associated with, e.g., matter flows''. In an ad hoc fashion, this noise is modelled as multiplicative (i.e.\ proportional to the density) Gaussian white noise with a time dependent amplitude that follows a negative power law in the expansion rate. The latter ensures that the noise term becomes small in the past (relative to the other terms in the continuity equations). The `average' parameters of the Universe are then obtained by ensemble average over the patches.

This construction leads to several results that seem baffling at first. For example, \citet{Lapi:2025mgr} find that ``the noise also alters the matter equation of state, which starts with the usual $w_m \approx 0$ at early times (i.e., pressureless dust), but progressively lowers toward $w_m \leq -1/3$ at late times''. Overall, the ensemble-average expansion rate and densities do not follow the standard Friedmann evolution at all, giving rise to accelerated expansion (without a cosmological constant) and drastically reduced dilution behaviour of matter and radiation under volume change.

Lapi et al.\ proceed undeterred and produce fits to cosmological observations, suggesting that their model may, at least partly, address the coincidence problem as well as the Hubble and $f \sigma_8$ tensions, among other things. However, even before we get to this stage, I think there are serious problems with their approach.

\newpage

\section{Problems}
\label{sec:problems}

\subsection{The need for a stochastic description}

Although it may seem superficially motivated, the need for a stochastic description at scales of tens of Mpc is actually poorly justified. On these mildly nonlinear scales, the randomness is entirely encoded in the initial conditions while the evolution remains completely deterministic and predictable. The dynamical timescale of structures of such scale is of the order of gigayears, meaning that the evolution can be robustly solved numerically over the entire age of the Universe, as is done, for example, in N-body simulations. In other words, the coarse-grained fluctuations that Lapi et al.\ want to model are not the result of a stochastic process that injects randomness over time, but rather the deterministic outcome of fluctuations that were there from the very beginning and whose properties have been well characterised.

\subsection{The noise modelling}

Given the previous point, if one were to pursue the stochastic description, it would appear that the noise properties could be derived from first principles, e.g.\ by measuring them from N-body simulations. However, the authors instead postulate some specific noise parameterisation and then obtain the parameters by fitting to observations. The link to the underlying physics is completely lost in this approach. Importantly, the finding of an accelerated expansion becomes circular and it is, at best, unclear if this is a physically meaningful outcome. In many ways, the approach is similar to fitting a generic dark energy parameterisation like $w_0$, $w_a$ \citep{Chevallier:2000qy,Linder:2002et}, without further explanation of how matter can behave in this way. The issue is not phenomenological modelling per se. Parameterisations such as $w_0$, $w_a$ are explicitly agnostic about microphysics. By contrast, $\eta$CDM claims to derive its behaviour from the standard gravitational dynamics of classical matter alone and, therefore, must remain consistent with the dynamical constraints of that system for which we already have an extremely detailed understanding.

\subsection{Using ensemble averages}

Maybe the most important problem of the approach is the inappropriate use of ensemble averages. In short, the framework assumes without justification that any background quantities, such as expansion history or density parameters, that enter the computation of observables, such as the distance-redshift relation or the growth rate, should be replaced by the ensemble average over the patches. That this can lead to rather unphysical results is maybe most clearly demonstrated considering the example of ensemble-averaged density, as used extensively in the present framework.

To demonstrate the point, let us assume that the Universe is divided in an equal number of patches that are initially either slightly under-dense or slightly over-dense. Let us further assume that all the over-dense patches eventually form some gravitationally bound structures that essentially stop evolving in density, while all the under-dense patches become voids that keep expanding indefinitely. It is easy to see that the ensemble average of the density, computed as $\left\langle \rho\right\rangle = N^{-1} \sum_{i=1}^N \rho_i$, then tends to a constant\footnote{This behaviour is similar to the attractor solution found in \cite{Lapi:2023plb,Lapi:2025mgr}, yet for entirely different reasons.} which corresponds to half the typical density of a collapsed patch. This average has of course nothing to do with what we mean by the average density of the Universe, which in this scenario would be computed as $\bar{\rho} = \sum_{i=1}^N (V_i \rho_i) / \sum_{i=1}^N V_i$, properly keeping track of the volume.

What the framework fails to establish is a clear justification why the ensemble average should be a useful tool for describing the observed Universe in the first place. The fallacy is sometimes a bit more subtle in other back-reaction frameworks such as AvERA \citep{Racz:2016rss} and the silent universe approximation \citep{Bolejko:2017wfy}, but at closer inspection they often also employ ensemble averages in a poorly justified manner. The problem with using ensemble averages in that way is that they introduce an operational definition of cosmological parameters that is conceptually divorced from the actually relevant operational definition used in observations.

To further elaborate on this point, let us now turn to some possible operational definitions of the Hubble parameter $H_0$. In a homogeneous FLRW model, $H_0$ specifies the local expansion rate. One can imagine observing a Hubble diagram, accepting for the sake of the argument that standard candles exist in a perfectly homogeneous Universe. $H_0$ is then also a parameter that describes the distribution of the observed data in this diagram. Furthermore, one can imagine dividing the spacetime into patches and computing $H_0$ via the ensemble average of the local expansion rate found in each patch. In the stated homogeneous limit, both operational definitions give the same answer.

Let us now consider the inhomogeneous case. From observations of standard candles, we can still measure a Hubble diagram. $H_0$ still operationally remains a parameter of the underlying distribution from which the data are drawn. Using forward modelling, we can, in principle, predict this underlying distribution as a function of $H_0$ and other parameters. Of course, the whole idea of cosmological analysis exploits this fact to derive credible `posteriors' for $H_0$, typically jointly with other parameters of the forward model. This means that the relationship between $H_0$ and the data is fundamentally maintained through our forward model, to the extent that this relationship \textit{defines} the meaning of $H_0$. Any alternative notions of $H_0$ are irrelevant from the perspective of cosmological parameter inference.

In particular, computing the ensemble average of the local expansion rate of patches has no obvious link to the observed Hubble diagram. Proponents of the use of such ensemble averages might argue that they are just constructing a more faithful forward model, accounting for inhomogeneities in the matter distribution. But this seems to be double-counting their effect since the standard forward model already predicts the distribution of data in the fully perturbed Universe. Case in point, full numerical simulations, the kind of which are routinely used in forward models, already account for the gravitational dynamics from horizon scales down to scales way below 10 Mpc. This brings me to my next point.

\subsection{Disregarding empirical evidence}

At scales of 10 Mpc and larger, we are clearly in the regime where we can make robust predictions from first principles, i.e.\ without invoking any effective description. One could argue that this regime is even quite accessible to analytical methods like Lagrangian perturbation theory, but it can certainly be studied numerically with exquisite accuracy, and indeed has been in countless independent studies that collectively yield a very consistent picture. Returning to the observed Hubble diagram, it is well understood that peculiar matter flows (ostensibly the source of the noise term in $\eta$CDM) introduce a correlated scatter in the observed data at low redshift that becomes less and less important at high redshift \citep{Hui:2005nm}. At high redshift, weak lensing magnification skews the distribution of the data \citep{Holz:2004xx}. Both effects are clearly within the error budget of current measurements and do not lead to large biases \citep{Fleury:2016fda}.

Strong empirical evidence is provided by full numerical forward models. \citet{Breton:2020puw} construct a Hubble diagram from a high-resolution cosmological N-body simulation, employing a ray-tracing method that accounts for all relevant effects of inhomogeneities. Their detailed study confirms the theoretical expectations about the distribution of the observed data. Their simulation is based on Newtonian gravity, but the conclusions should be valid with regard to $\eta$CDM, as it is nowhere claimed that the phenomenology of $\eta$CDM somehow relies on general relativistic corrections to the Newtonian picture.

However, if a fully relativistic forward model is desired, examples have been presented in \citet{Adamek:2018rru} and \citet{Macpherson:2022eve}. In both cases, structure formation is simulated in a general relativistic spacetime, and relativistic ray tracing is employed to construct the synthetic observations. The two approaches are completely independent and use different methods, yet both find a scatter of the data consistent with expectations and no significant modification of the inferred expansion history. Given that these works fully account for possible back-reaction in the evolution of spacetime \textit{and} all kinematical effects in the construction of observables on the light cone, they present the strongest empirical evidence against claims of large (order unity) effects of cosmological back-reaction \citep[see also][]{Oestreicher:2024mtw}.

It should be noted that \citet{Lapi:2025mgr} are well aware of these results, citing the related works of \citet{Adamek:2017mzb} and \citet{Macpherson:2018btl}. They dismiss the evidence with following argument: ``However, these conclusions strongly rely on the assumption that it is possible to define a global background metric (or a global rescaling of it to different scales) and a deterministic backreaction term. This is not at all guaranteed at late times on scales $\lesssim 50\,h^{-1}\,\mathrm{Mpc}$ pertaining to the cosmic web and relevant for several cosmological observables. There the different evolutions of the various patches due to local inhomogeneities/anisotropies, cosmic flows, tidal forces, sample variance, and many complex gravitational processes cannot be neglected and call for a statistical approach to cosmology as pursued in the main text''. This argument is misplaced for several reasons. First, with regard to making their own predictions for $\eta$CDM, \citet{Lapi:2025mgr} state elsewhere in the text: ``Since the inhomogeneities/anisotropies described by the noise term are expected to be small, it is a reasonable approximation to assume that the ensemble-averaged Universe is described by a Friedman-Robertson-Walker metric, and to adopt the standard redshift-distance relation''. It is hard to see how both arguments can be valid at the same time, and it would appear that their own analysis \textit{explicitly assumes} that it is indeed possible to define a global background metric. This shows that the first part of the argument is spurious.

Second, still on the possibility of defining a global background metric, this is not actually an assumption, but rather an empirical result of the studies by \citet{Adamek:2017mzb} and \citet{Macpherson:2018btl}. Both studies explicitly construct valid spacetime solutions that are faithful solutions of Einstein's equations with inhomogeneous matter. The fact that the metric can be written as the Friedmann--Lema\^itre--Robertson--Walker metric with small fluctuations is not an assumption here but the result of a direct calculation.

Third, concerning the part about ``deterministic backreaction'', the claim seems to be that deterministic evolution is not guaranteed at scales $\lesssim 50\,h^{-1}\,\mathrm{Mpc}$. As noted earlier, this position is untenable. Clearly, the underlying theory of gravity is deterministic in nature. There is the possibility of deterministic chaos in the sense that small changes in initial conditions lead to exponentially divergent solutions. But within the age of the Universe this only happens on much smaller scales than the quoted $50\,h^{-1}\,\mathrm{Mpc}$ scale. This is obvious from the dynamical timescale of gravitational collapse on those scales. Furthermore, for the purpose of modelling the observed data, we mostly do not even care about the precise orbits of every matter element, but only about their statistical properties. This is because we want to predict the distributions from which our data are drawn rather than the specific realisations. These can be obtained robustly and deterministically from our forward models, and they explicitly include the variance coming from uncertainty on the initial data.

Finally, in the last sentence of their misplaced argument, \citet{Lapi:2025mgr} raise the point that different evolutions of various patches due to inhomogeneities/anisotropies, cosmic flows, etc. cannot be neglected. Certainly, this is not a valid criticism against \citet{Adamek:2017mzb} and \citet{Macpherson:2018btl}, since their simulations (and simulations of structure formation in general) are designed \textit{precisely} to capture these effects.

\newpage

\section{Discussion}
\label{sec:concl}

I have argued that the $\eta$CDM model, proposed by \citet{Lapi:2023plb,Lapi:2025mgr} as an alternative to $\Lambda$CDM, does not withstand critical examination. I highlighted that several pillars of the model are poorly conceived and that the claimed phenomenology contradicts established empirical evidence. Perhaps the fundamental issue is that the authors are trying to avoid introducing any new physics, instead hoping to discover new dynamics by merely changing the methodology used to describe the \textit{same} physical system. In doing so, they must confront the reality that all established descriptions, including extremely robust and detailed forward models, fail to corroborate their results. While we should always be prepared to critically examine our established methods, the same must be true of any methods we seek to replace them with. On balance, I would argue that this is where ``Little ado about everything'' falls short. Although the desire to challenge $\Lambda$CDM and solve the outstanding tensions is understandable, we should not tempt ourselves to compromise on scientific rigour along the way.

\section*{Acknowledgments}

I thank A.\ Lapi, L.\ Boco, M.\ M.\ Cueli, B.\ S.\ Haridasu, T.\ Ronconi, C.\ Baccigalupi and L.\ Danese for their hospitality during a memorable visit to SISSA/Trieste in April 2024 where we had the opportunity to discuss their ideas on $\eta$CDM at great length. I also thank P.\ Fleury and an anonymous referee for valuable comments. My work is partly funded by the Swiss National Science Foundation and by the Dr.\ Tomalla Foundation for Gravity Research.

\bibliographystyle{mnras}
\bibliography{comment}

\end{document}